\documentclass[preprint,singlespace]{icarus}
\usepackage{graphicx}
\usepackage{geometry}
\usepackage{times}
\singlespace

\begin{document}

\begin{center}

Title of the paper:

The shape distribution of asteroid families -- evidence for
evolution driven by small impacts

\bigbreak

Authors:

Gyula M. Szab\'o$^{1,2\ast}$, L\'aszl\'o L. Kiss$^3$

\normalsize{$^{1}$Magyary Fellow, School of Physics A28, University of Sydney, NSW 2006, Australia,}\\
\normalsize{$^{2}$Department of Experimental Physics, University of Szeged, 6720 Szeged, Hungary,}\\
\normalsize{$^{3}$School of Physics A28, University of Sydney, NSW 2006, Australia}\\
\normalsize{$^\ast$E-mail:  szgy@titan.physx.u-szeged.hu}

%\newpage
\vskip2cm

Proposed runnung head:

The shape distribution of asteroid families

\bigbreak

Name and address to which editorial correspondence and proofs should be
directed:

Gyula M. Szab\'o

University of Szeged

Dept. Experimental Physics

D\'om t\'er 9.

H-6720 Szeged, Hungary

e-mail: szgy@titan.physx.u-szeged.hu

\bigskip

\newpage

Abstract

A statistical analysis of brightness variability of
asteroids reveals  how their shapes evolve from elongated to rough spheroidal
forms, presumably driven by  impact-related phenomena. Based on the Sloan
Digital Sky Survey Moving Object Catalog, we determined the  shape distribution
of 11,735 asteroids, with special emphasis on eight  prominent asteroid
families. In young families,  asteroids have a wide range of shape elongations,
implying frag\-mentation-formation. In older  families we see an increasing
number of rough spheroids, in agreement with the predictions of an
impact-driven evolution. Old families also contain a group of moderately
elongated members, which we suggest correspond to higher-density, more
impact-resistant cores of former fragmented asteroids that have undergone slow
shape erosion. A few percent of asteroids have  very elongated shapes, and can
either be young fragments or tidally reshaped bodies. Our results confirm that
the  majority of asteroids are gravitationally bound ``rubble piles''.


\end{center}

\section{Introduction}

About one third of all known asteroids belong to families (\cite{zappala}), which
are clusters of asteroids believed to result from collisional disruption of
parent bodies (\cite{brien}).  The hypothesis of collisional origin is based entirely on the
observed similarity of dynamical and spectral properties, and not on the
understanding of the collisional physics itself (\cite{michel2001}). Members of
asteroid families share similar orbital elements (semimajor axis,  eccentricity
and orbital inclination) (\cite{nesvorny+}) and chemical composition
(\cite{juric}). They often have low density { which suggests a high porosity
-- the asteroids are often made of} gravitationally bound fragments or 
``rubble piles'' (\cite{richardson}). { This structure is consistent with a
collisional origin, and} explains  both the lack of fast rotators among
asteroids larger than a few hundred meters (\cite{pravec}) and the presence of
large craters revealed by close-up surface images (\cite{chapman}). The
rotational rates of the asteroids display a Maxwellian distribution, indicating
angular momentum redistribution by collisions (\cite{binzel,fulchignoni}). As
such, prominent asteroid families are therefore important tracers of
high-velocity impacts, one of the principal geologic processes affecting small
bodies in the Solar System.

Fig 1. comes here

Collisions modify the shape and size of an asteroid. { The most energetic
collisions disrupt the bodies and generate a number of smaller, gravitationally
unbound fragments which will evolve further as new individual asteroids. The
less energetic collisions can only fragment the body and/or form craters,
without disruption. Collisions are rare, and therefore cannot be observed
directly, and their complete understanding requires numerical  experiments that
have to reproduce the observable constraints of the impacts. Currently these
constraints  are: asteroid size distribution, the number of asteroid families,
the age distribution of meteorites, and surface cratering of 5 asteroids which
were encountered by spacecraft. The last of these cannot be easily interpreted,
because craters can overlap or be completely saturated, and
crater size depends on the impactor/crater scaling law model. The other
constraints are only sensitive to the most energetic, disruptive collisions.
Significant information on the less energetic impacts may be extracted from
asteroid shapes: it
has been proposed  (\cite{richardson,K+A}) that the shape of asteroids can evolve
with age due to several microimpacts. However, there are only
a few asteroids with known shape and hence the evolutionary effects could not
have not been fully exploited yet.}

While sizes can be constrained relatively easily from absolute photometry
(\cite{eddington}), determining shapes is much more difficult. Space missions
have visited only a few asteroids and shapes for these have been determined
accurately (\cite{fujiwara,abe,okada,saito}). Radar observations have revealed
shapes of  Near-Earth Objects encountering the Earth ( e.g. \cite{ostro}), while
high-resolution imaging is restricted to the largest asteroids
(\cite{tanga,marchis}). For the rest of the asteroid belt, time-resolved
photometry is the only method to determine the shape. The key effect is the
rotation: it causes periodic brightness variations which, in principle, reveal
the shape of a given asteroid if there are enough data for full inversion of
the lightcurve, or at least for fitting a triaxial ellipsoid.  Despite the
relatively simple principles of the method, there has been slow progress
because of the time-consuming nature of the observations.  Recent advances in
all-sky surveys offer new approaches to the problem. These surveys often
observe thousands of asteroids in a night,  and with the appropriate data
processing the shape distribution can be  constrained. Here we deduce the shape
distribution of 11,735 main-belt asteroids belonging to eight prominent
asteroid families, using data from the Sloan Digital Sky Survey
(SDSS).

SDSS was a deep imaging survey covering over a quarter of the Celestial Sphere
in the Northen Galactic Cap, resulting in high-quality photometric measurements
of 50 million stars and a similar number of galaxies. Although its main purpose
was cosmology, SDSS has also collected  204,305 measurements of asteroids,
listed in the SDSS Moving Object Catalog 3 (SDSS MOC) \footnote{Available at
http://www.sdss.org} (\cite{ivezic}).  Of these, 67,637 data points have been
linked to 26,847 known asteroids (\cite{juric}). This means that all the
identified asteroids have been observed an average of 2.5 times. The
calibration of the SDSS is very accurate because all measurements were
acquired with the same instrument and from the same observing site. Therefore, 
systematic effects are minimized and the individual brightness measurements are
accurate to 1--2\%{} down to magnitude 20 in $r$ band.

\section{Data and Analysis}

We selected eight asteroid families for a detailed analysis of the shape
distribution and its possible evolution. We chose families with at least 400
pairs of measurements from SDSS that also have a published age determination
(\cite{nesvorny+,carruba}).  This is satisfied for five families (Flora, Vesta,
Eos, Eunomia, Themis). We also included two very old families (Koronis, Hygiea)
and one quite young (Massalia), each with fewer than 400 pairs of observation,
which helped expand the analysis to a broader range of ages. We selected
the family members according to their orbital elements (\cite{szabo}). The
selection criteria and the number of members are summarized in Table 1, while
their positions in the orbital semi-major axis -- orbital inclination plane are
plotted in Fig. 1.

{  Brightness variations are caused by rotation of a non-spherical body  and also
by changing geometric conditions (the Sun-asteroid-Earth angle and the
Sun-asteroid and asteroid-Earth distances). To reduce the geometric effects, we
restricted our sample to those  asteroids for which the geometric
conditions of the two independent  observations were very similar.}

The selection steps from SDSS MOC were as follows. 
In order to minimize the effect of variable angle from the opposition, 
we selected only those objects for which the change of this angle was less than 2.5 degrees. 
Similarly, because the variation of the solar phase affects the brightness of the asteroid,
we required this to change by less than 1.5 degree change in it.
In order to guarantee that rotational phases were uncorrelated, only pairs of observations
more than 1 day apart were considered. The typical rotational periods of asteroids fall between
0.3 d and 0.6 d and the lightcurve regularly has two maxima and two minima. Consequently, our smallest 
sampling interval is comparable or longer than one hump on the light curve in virtually all
cases, resulting in uncorrelated sampling.
In order to avoid the increased photometric errors at the faint end, we 
required $m < 20$ in the $r$ photometric band.

{  
This way we
ended up with 11,735 pairs of  photometric observations.  The resulting
cumulative distribution of the measured brightness variations is plotted with
the thick line in Fig.\ 2A. We see that for about 80\% of the asteroids the
difference between the two measurements was less than 0.2 mag, but that some
large differences of up to 0.8 mag, indicating highly elongated shapes.

We tested that the chosen magnitude limit of $r<20$ is not too faint by checking whether fainter asteroids
in our sample
showed apparently larger brightness variations. A positive result would have implied that either the
selection limit was not strict enough or the smaller asteroids were more elongated in the examined
size range. We found the contrary: using a Kolmogorov-Smirnov test we compared
the brightness variations of bright ($<$ 18.5 magnitude) and faint (between 18 and 20 magnitude)
objects, and found them to be indistinguishable.
This shows both the high quality of the data for the faintest asteroids in the sample
and the independence of the average shape elongation on size in the studied 1--30 km diameter range.}

As the brightness depends on the Sun-asteroid and the asteroid-Earth distances,
the measurements were converted to absolute magnitudes as:
\begin{equation}
	\Delta m_{1,2}^{obs}=| m_1-5\log(D_1 R_1) - m_2 + 5\log(D_2 R_2) |,
\end{equation}
Here $D_1$, $R_1$ and $D_2$, $R_2$ are the geocentric ($D$) and heliocentric 
($R$) distances of the asteroid in AU, at the epochs of the two observations. 
We used $r$-band magnitudes because they are the most accurate in the color range of asteroids.
The outlined procedure reduced the geometrical 
effects below $\approx0.02$ magnitudes in $\Delta m_{1,2}^{obs}$.

Table 1. comes here

\section{Methods of fitting}

{  If we know only the brightness difference displayed by an asteroid 
between two epochs, the exact shape elongation cannot be determined. 
However, the distribution of the shape elongation can be determined for a 
large ensemble of asteroids with a statistical inversion. 
We have developed this }method to determine the shape distribution from
sparsely sampled observations of many asteroids.  The basic idea is to use the
two-point statistics of photometry to determine the shape elongation, which we
parametrize by the shape axis ratio, $a/b$.  

{The light variations of asteroids are due to the departures from a spherical symmetry,
primarily the elongation of the shape. Rotation causes periodic variation 
of the illuminated surface, producing measurable brightness changes. 
The photometric amplitude reflects the ratio of the largest and smallest
projected area observed during a rotation, which 
is approximately the $a/b$ axis ratio of the best-fit ellipsoid.

Assuming a triaxial ellipsoid with axes $a>b>c$ that is rotating around the $c$ axis, and given
the aspect angle $\vartheta$ (the angle between the line of sight
and the axis of rotation), the projection of the shape can be calculated (Connelly and Ostro, 1984).
Expressed in magnitudes, the brightness varies as
\begin{equation}
	m(\phi)= m_0 + 1.25 \log \left( {\sin^2 \vartheta \sin^2 \phi \over a^2} +
	{\sin^2 \vartheta \cos^2 \phi \over b^2}+
	{\cos^2 \vartheta \over c^2} \right),
	\end{equation}
where $\phi$ is the rotational phase and $m_0$ is the maximum brightness.

If an asteroid has been sparsely observed, there is no way to determine the rotational period --
consequently the rotational phases and their relation to each other also remain unknown.
From a large sample, however, we can deduce the shape distribution as follows. Let us consider brightness measurements
at two randomly chosen rotational phases $\phi_1$ and $\phi_2$, whose difference
\begin{equation}
	\Delta m_{1,2}=
	| m(\phi_1) - m(\phi_2) | 
	\end{equation}
will be the basis of our method.
In the case of a single pair of observations, one cannot find a direct relation between $\Delta m_{1,2}^{obs}$ and $a/b$ because of the unknown rotational phases. For a statistically large sample, however, 
the {\it distribution} of the observed brightness differences can be compared to model predictions. The idea is that 
spherical asteroids never cause large brightness change
while more elongated ones sometimes do. Thus, if $\phi_1$ and $\phi_2$ are independent and uniformly
distributed, the different values of $\Delta m_{1,2}^{obs}$ are predicted to have different probabilities. By comparing the
model predictions to observations, the distribution of $a/b$ can be constrained as follows.

The cumulative distribution function of the observations is
\begin{equation}
	\xi(\Delta m_{1,2}) =  {1\over {\rm number}_{all}} \ {\rm number}(\Delta m_{1,2}^{obs}<\Delta m_{1,2}).
\end{equation}
We compare this to the cumulative distribution function of the models, which was calculated from one million
simulated observations by putting random observational phases in Eqs. 2-3.

Fig. 2. comes here
}

{With the help of {  these template simulations}, we reconstructed the observed 
brightness variations as the linear combination of {models for differently elongated bodies.} 
We calculated {  the template cumulative distribution of brightness variations} for 17 values of 
$a/b \equiv$ 1.1, 1.2, 1.3, 1.4, 1.5, 1.6, 1.7, 1.8, 1.9, 2.0, 2.2, 2.4, 2.6,
2.8,  3.0, 3.5, 4.0, each taken from $10^6$ model observations from a 
Monte Carlo experiment. 
{  We simulated $17\times 10^6$ observations with independent and uniformly distributed rotational phases. 
The observed cumulative distribution  was calculated using Eq. 4.}

Each thin line in Fig. 2A
shows the expected distribution of magnitude differences for a populations of
asteroids having a single value of $a/b$.

A comparison of the thick line and the thin lines in Fig. 2A shows that  the
observed brightness changes came from a mixture of asteroid shapes. We 
therefore fitted a linear combination of the templates to the measured curve. 
For this we applied a non-negative least squares method (\cite{lawson}).  The
best fit coefficients directly give the $a/b$  shape distribution responsible
for the observed variability. This can simply be  interpreted as the most
probable shape distribution in a family. { The uncertainties of each coefficients 
are less than $\pm3 \%{}
$
in all cases and in all $a/b$ bins.}  We plot the shape distribution for
the whole sample with the black solid line in Fig. 2B.

We tested this result using asteroid shapes estimated from full rotational
lightcurves. As of writing this paper, there are 1,207 asteroids in the
literature for which  multi-epoch full lightcurves were published in the past
70 years (\cite{kryszczynska}).  For each asteroid, the maximum amplitude can be
easily converted to $a/b$ shape  elongation. The maximal observed amplitude corresponds to the perpendicular
aspect angle of the rotation axis, therefore its value is approximately the $a/b$ axis ratio of the best-fit ellipsoid.
We plot the
resulting distribution with the {  orange (grey in BW)} shaded histogram  in Fig. 2B. The two
curves shown are completely independent and their excellent agreement confirms
the reliability of our method, which opens a new  way forward to determine
asteroid shapes on a massive scale.

Fig 3. comes here

{  Before turning to the results, we shortly discuss the
two-point brightness variation functions for the 8 asteroid families (Fig 3). 
In the upper panels we plot the cumulative distributions, while
the lower panels show the {\it difference} of the observed functions with respect to
that of the whole sample (Fig. 2, thick line). 
Here the observational data are shown by different symbols for different
families (see the labels) while the smooth lines represent the fitted models.

Visual inspection shows that some asteroid families (such as Massalia) contain
many elongated asteroids which display large amplitude two-point
brightness variations on average. This is indicated by the 
downward dip in their cumulative distribution
around 0\fm{}1--0\fm{}2, showing that a larger fraction of asteroids in this family
exhibited variations over a few tenth of a magnitude. Similarly, in other
families (such as the Vestas) the asteroids are more rounded in shape,
which is indicated by the higher rate of small ($<0.2$ mag) 
variations.
}

\section{Shape distribution of asteroid families}

Fig. 4. comes here

The shape distributions for the eight selected families are plotted in Fig. 4.
In order to separate age and distance effects,  the families have been grouped
in two sequences. The Massalia-Koronis sequence (Fig 3 A--D)  shows families
with a wide range of ages  (Massalia: 150 million years to Koronis: 2500
million years). The Vesta--Themis sequence (Fig 3 E--H)  contains very old
families at different  heliocentric distances. 

{ An important model parameter is the distribution of $\vartheta$, which is
related to the average pole inclination of the model asteroids. First we
assumed that all asteroids have a rotation axis which is perpendicular to the
line of sight, $\vartheta=90$ degrees.
However, the choice of the average inclination does
not greatly affect the interpretation. In order to show this, we present here a
second solution which takes the different inclinations into account. In this
calculation, the ecliptic longitude of the pole was randomly oriented, but the
latitude was fixed to be 50 degrees, because the average spin axis inclination
is 50 degrees for 104 asteroids with known pole orientation listed by 
(\cite{Magnusson2005}). The number of highly elongated objects have increased,
but in general, results are very similar to those given by the first
calculation (Fig. \ref{appfig}).}

Fig. 5. comes here

The shape distribution of the Massalia family confirms that young families
display a  variety of shapes. The mean $a/b$ is 1.39, significantly larger than
that of  the mean of the full sample and very close to the value of $a/b=1.41$
obtained from the laboratory  experiments with catastrophic collisions of
monolithic targets (\cite{cappacioni,catullo,ryan}). In the Massalia--Koronis
sequence, the fraction of spheroids ($a/b<1.2$) rises prominently toward the
older  families, finally exceeding 50\%{} in the old Koronis family. 

The discussed asteroid families have various orbital and
structure properties (size distribution, concentration etc.) which are not correlated with age. 
So the age dependent variations in the shape distribution are {\it not}
due to the variation of orbital or structural parameters. 
We interpret this marked change in the shape distribution with an evolutionary
effect: the young bodies, whose shapes are presumably determined
by collisional disruption, are reshaped into more rounded forms in 1--2 billion
years.

In order to check this conclusion again, we examined possible correlations between
the distribution of the observed brightness variation and the eccentricity, inclination, solar phase at the epoch of observation
and the size of asteroids, while no correlations have been found. This proves that the variation of the shape distribution is
age dependent, i.e. an evolutional outcome.

Fig. 6. comes here

{  Several families in Fig. 4 show bimodal distribution of the axis
ratios. Being a fundamental result, we test its robustness.
For instance, possible photometric problems in the dataset could result in
very large but spurious variation, which might then lead to a second or third
peak in the $a/b$ distribution. 
Following a Referee's suggestion, we clipped the large
values from the cumulative two-point brightness variation (all measurements
having $\Delta m_{1,2}>0.65$ were ignored) and repeated the determination of
the $a/b$ distribution. The results (Fig 6) confirm the presence of the second
peak.}

We suggest this evolution to be due to impact-driven reshaping and
impact-induced seismic activity erosion. After the families were formed, the
asteroids have been eroded by continuous impact cratering and possibly further
disruptive collisions.  The impact origin theory of shape evolution
(\cite{leinhardt}) argues that subcatastrophic impacts (an excavation followed by
the launch, orbit and reimpact of ejecta) rearrange the regolith layer of the
colliding bodies. This process has had enough time to reshape the older
asteroids by now. Numerical simulations of this scenario \cite{K+A} revealed
that both oblate and cigar-like elongated prolate shapes  evolve to become
close to oblate spherical after about 10,000 impacts.  The more rapidly the
body was spinning, the more flattened the outcome, with $a/b$ between 1.0--1.35
in all calculations.  The impact-induced seismic activity theory
(\cite{richardson}) predicts that subcatastrophic impacts generate seismic
shaking in the target asteroid, and the regolith layer moves  then downslope.
In case of an Eros-like, 1.5 km sized asteroid, covered by a  layer of regolith
tens of meter thick, 0.5--10-meter-sized impactors are the best candidates for
such process. This process is in agreement with the surface structures and the
erased small craters on Eros (\cite{richardson,chapman2002}).

Both subcatastrophic impacts and impact-induced seismic activity require enough
time for reshaping  before the characteristic shape properties are destroyed by
a catastrophic collision. This assumes quite a steep power-law for the size
distribution of impactors, with only a few large bodies. Such a steep
population explains the crater population of some asteroids (\cite{greenberg})
and is also in agreement with the idea that large family-forming asteroids have
been battered and pre-shattered by softer impacts prior to the family-forming
disruption  (\cite{michel}). Our results lead to the conclusion that shapes
indeed evolve toward spherical symmetry, in agreement with the  predictions. We
suggest that impact reshaping and impact-induced seismic activity are dominant
processes in the late evolution of asteroid shapes.

{}

The Vesta--Themis (Fig 4. E--H) sequence shows that the erosion depends on the
semi-major axis of the orbit. These four families are quite old, with estimated
ages of about 2 billion years, while their heliocentric distances range from
2.35 AU to 3.2 AU. Our results imply that the erosion of shapes becomes slower
with the increasing heliocentric distance. This can be qualitatively explained
by the effect of smaller orbital velocities at large solar distances, which
lead to less energetic collisions with impactors. There may also be an effect
from different  chemical compositions (silicates and basalts close to the Sun,
chondrites farther out), since different materials can fragment differently
(\cite{K+A}).

Some of the shape distributions are multimodal (Eos, Koronis, Eunomia, Hygiea,
Themis), which might indicate different compositions and/or internal
structures. Among Near-Earth asteroids at least three types have been
identified: ``strengthless bodies'' with high porosity close to equilibrium
shapes (\cite{ostro}); ``potatoes'', which have  larger density and a somewhat
irregular shape; and irregular two-lobed ``dog bones'' or ``dumbbell''-shaped 
asteroids (\cite{ostro_bone}), which look like merged bodies.   Simulations have
shown that strengthless bodies do not have perfectly symmetrical shapes (mainly
because of the stochastic nature of cratering)  (\cite{K+A}), thus we may
identify the peaks at $a/b=$1.2 in the evolved families with the loosest
structure rubble piles.  The second clump at around $a/b=$1.5--1.7 may consist
of the higher-density fragments, either rubble piles or monolithic  structures,
which evolve more slowly toward the spherical shape.  There is a third
population of very elongated members in all families, which are similar to the
extreme examples of Geographos (\cite{ostro_geogr}) and Cerberus (\cite{cerbe}),
both belonging to the Earth Crossing Objects. Our results show that such
asteroids exist all over the Main Belt (see also the ``dog-bone'' like asteroid
Kleopatra \cite{ostro_bone}). It has been  suggested that the single-convex
side, elongated bodies such as 1620 Geographos and 433 Eros can be formed by
tidal forces during close encounters with Earth and Venus, and about 2\%{} of
the rubble piles are likely to suffer tidal shaping (\cite{bottke}).  While this
mechanism might explain the strong elongation in the Near-Earth population, it
is hardly applicable to the Main Belt. Indeed,  the shape of the Main Belt
asteroid Kleopatra was ascribed to an exotic sequence of collisional events
(\cite{ostro_bone}).  Our results shows that  very elongated asteroids ($a/b>$2)
make up a few percent, suggesting that those exotic collisional events may have
happened much more frequently than has been anticipated: up to 4--5 percent of
asteroids in certain families (Flora, Eos, Themis, Hygiea) have very elongated
shapes.

To conclude, the main results of the paper are as follows. We have
determined the shape distribution for almost 12,000 minor planets from pairs of
accurate brightness measurements using the SDSS database. The overall
distribution is in good agreement with that based on long-term photometric
monitoring but increases the sample by more than an order of magnitude. Our
large sample allows us to investigate differences between minor planet families
for the first time.  We have determined the empirical shape distribution
function for  eight large families, which are defined by their clumpiness in
the phase space of orbital elements and have estimated dynamical ages. We find
a tendency for old families to have significantly rounder members, which we
interpret as evidence for accumulated  shaping by low-energy impacts.

\begin{acknowledgements}
This work has been supported by 
the Magyary Zolt\'an Public Foundation for Higher Education, the Bolyai J\'anos Research Fellowship and a
University of Sydney Postdoctoral Research Fellowship.
\end{acknowledgements}

\newpage

\begin{table*}[h]
\label{Table1}
\caption{Selection volumes of asteroid families in the space of orbital elements. The number of identified objects and the
estimated ages are also given.}
\begin{tabular}{lllllll}
\hline
Families & $a_p$ & $\sin i$ & $e$ & $N$ & Age & Ref\\
 & AU & & & & Gyr\\
\hline
Flora & 2.16--2.32 & 0.04--0.125 & 0.10--0.18 & 819 & 0.5, 1.0 & \cite{carruba,nesvorny+}\\
Vesta & 2.28--2.41 & 0.10--0.135 & 0.07--0.13 & 482 & $>$1.3& \cite{carruba}\\
Massalia & 2.35--2-46 & 0.01--0.04 & 0.14--0.18 & 132 & 0.15& \cite{nesvorny+}\\
Eunomia & 2.52--2.72 & 0.19--0.26 & 0.12--0.19 & 824 & 2.5& \cite{nesvorny+}\\
Koronis &2.81--2.96 & 0.03--0.05 & 0.00--0.10 & 214 & 2.5& \cite{nesvorny+}\\
Eos & 2.95--3.10 & 0.15--0.20 & 0.04--0.11 & 540 & 1.3& \cite{nesvorny+}\\
Themis & 3.03--3.23 & 0--0.6 & 0.11--0.20 & 968 & 2.5& \cite{nesvorny+}\\
Hygiea & 3.08--3.23 & 0.07--0.11 & 0.10--0.16 & 193 & 2.0 & \cite{nesvorny+}\\  
\hline
\end{tabular}
\end{table*}

\newpage
Figure captions

\bigbreak

Fig. 1.
{The orbits of 26,847 asteroids in the SDSS MOC 3 catalog. Families studied in this paper are labeled.}

Fig. 2. {A. the observed cumulative distribution of brightness differences of 11,735 SDSS MOC asteroids (thick line) 
	with the model templates for discrete $a/b\equiv$1.1,1.2,1.3... shape elongations (thin lines). 
	B: the shape distribution for all the 11,735 
	asteroids determined by our method (thick line) and the histogram of shapes 
	from 1207 published lightcurves (shaded area). The excellent agreement confirms 
	the reliability of the statistical approach.}

Fig. 3. {Upper panels: the cumulative brightness variation of the examined
	families; lower panels: the difference in respect to the cumulative brightness
	variation in the whole sample. Individual symbols show the observational data,
	smooth lines represent the fitted models. The two group of families
	are the same as in Fig. 4.}

Fig. 4. {The evolution of shape distribution in asteroid families. A-D: (Massalia to 
	Koronis) families of 150 to 2500 million years show the dependence on age: 
	the more elongated members of young families erode in time toward rough 
	spheroids, whose relative frequency increases up to 
	$50\%
	$. E-H: (Vesta to 
	Themis) old families at increasing heliocentric distances -- the farther the 
	family, the more elongated its members. Note the distinct peaks of the 
	distributions.}
	
Fig. 5. {The same as Fig 4, but calculated with model asteroids having randomly orinted spin axis with a fixed latitude of 50 degrees.}

Fig. 6. {The same as Fig 4, but ignoring asteroids with $\Delta m_{1,2}>0\fm{}65$, with which we checked
	the robustness of the bimodal distributions.}

\newpage

\begin{figure}[h]
\begin{center}
\centering\includegraphics[width=8.5cm]{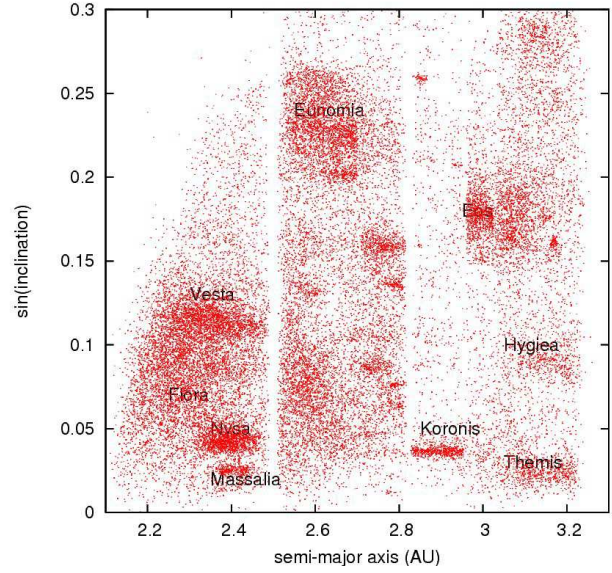}
\caption{The orbits of 26,847 asteroids in the SDSS MOC 3 catalog. Families studied in this paper are labeled.}
\label{distr}
\end{center}
\end{figure}

\newpage

\begin{figure}[h]
\begin{center}
	\centering\includegraphics[width=11cm]{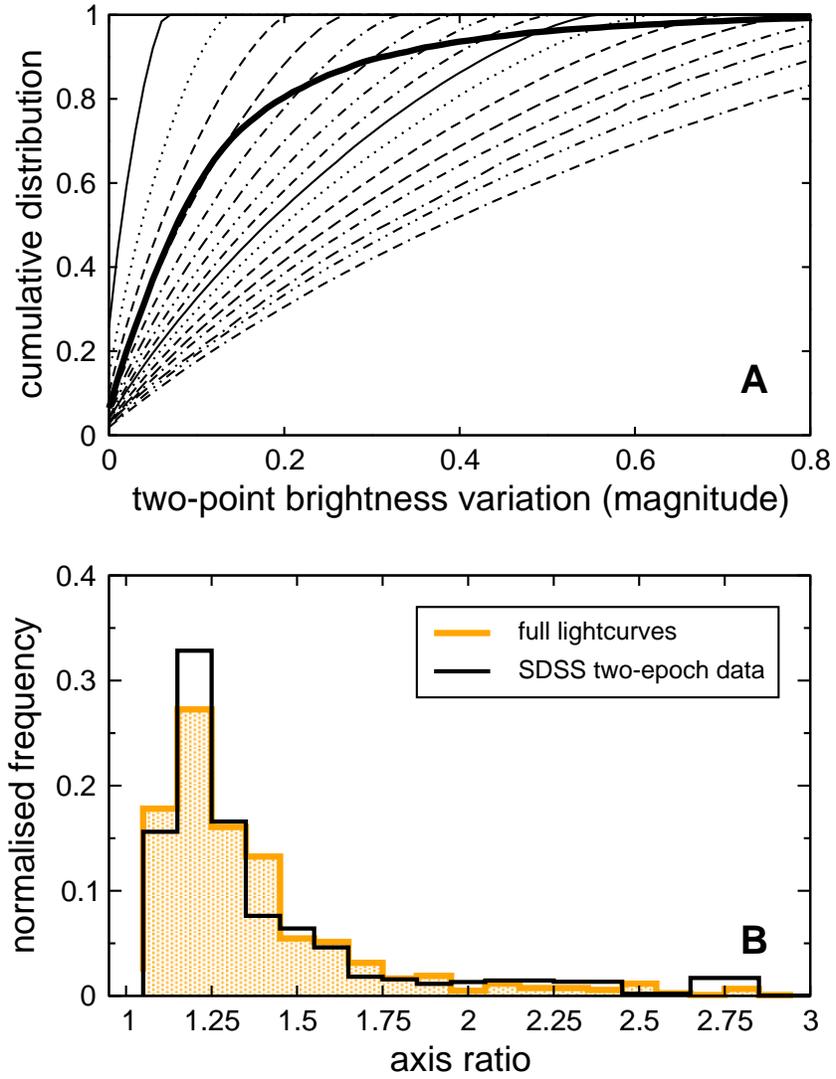}
	\caption{A. the observed cumulative distribution of brightness differences of 11,735 SDSS MOC asteroids (thick line) 
	with the model templates for discrete $a/b\equiv$1.1,1.2,1.3... shape elongations (thin lines). 
	B: the shape distribution for all the 11,735 
	asteroids determined by our method (thick line) and the histogram of shapes 
	from 1207 published lightcurves (shaded area). The excellent agreement confirms 
	the reliability of the statistical approach.}
	\label{fitfig}
\end{center}
\end{figure}

\newpage

\begin{figure}[h]
	\includegraphics[width=8cm]{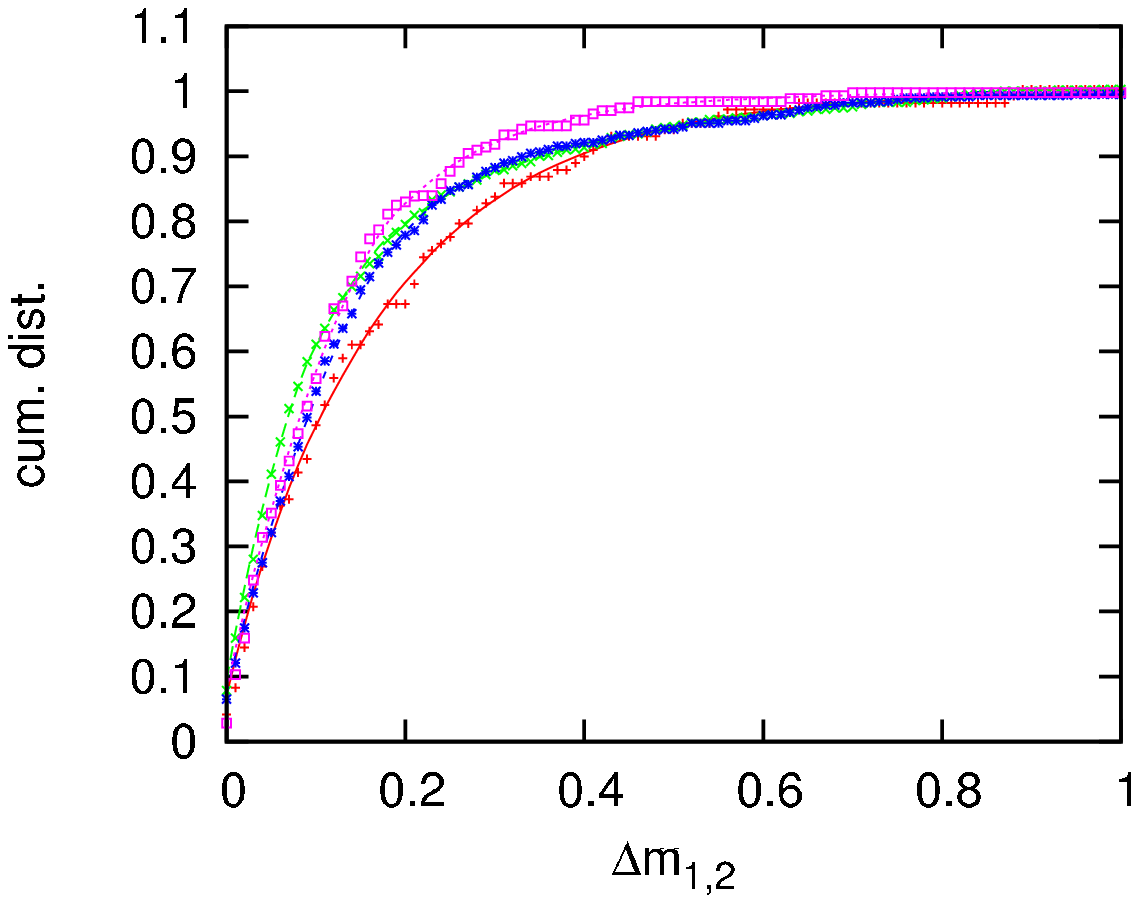}
	\includegraphics[width=8cm]{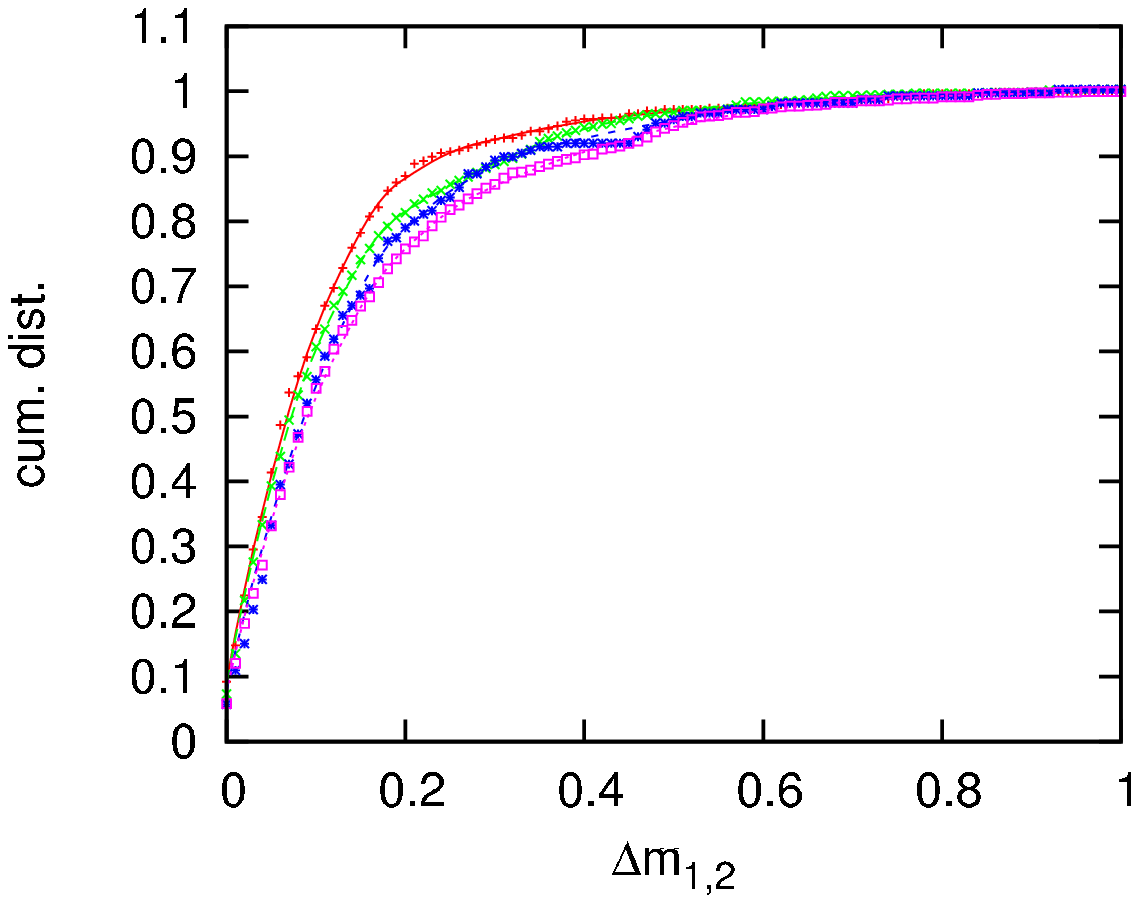}
	\includegraphics[width=8cm]{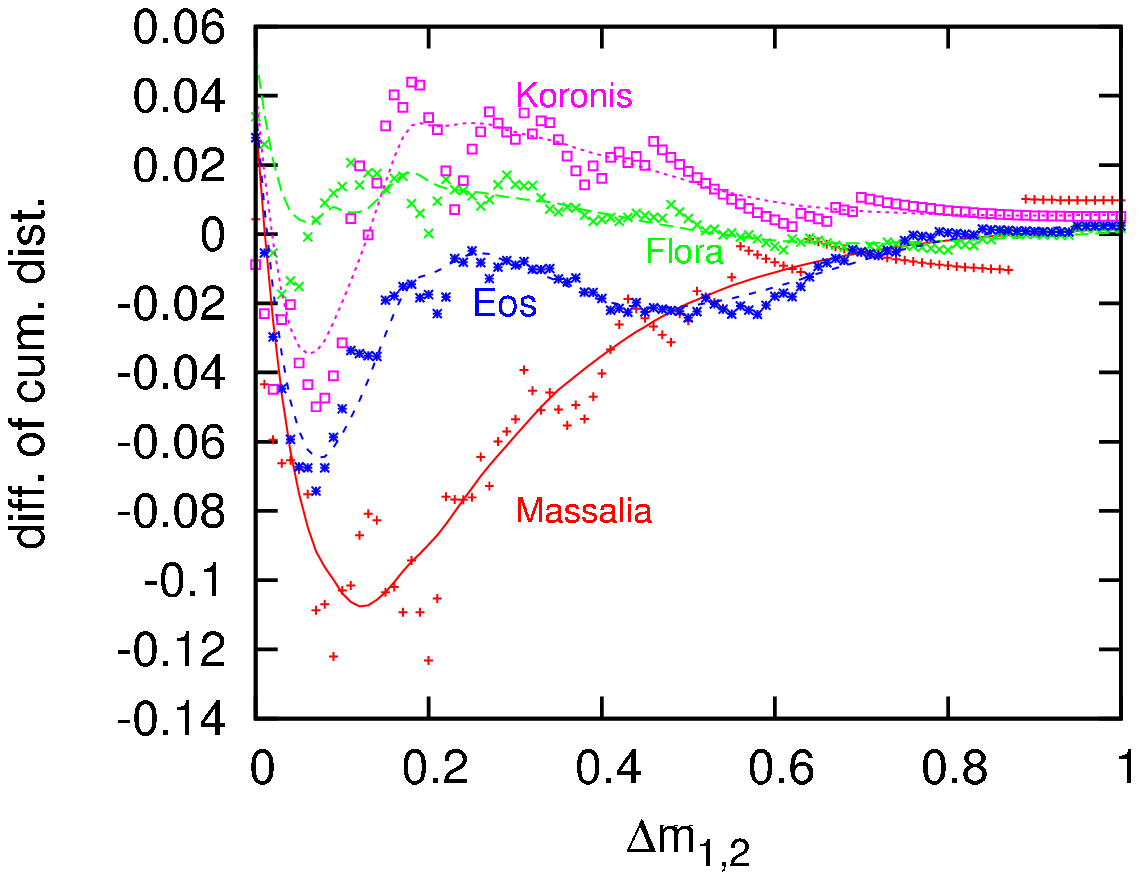}
	\includegraphics[width=8cm]{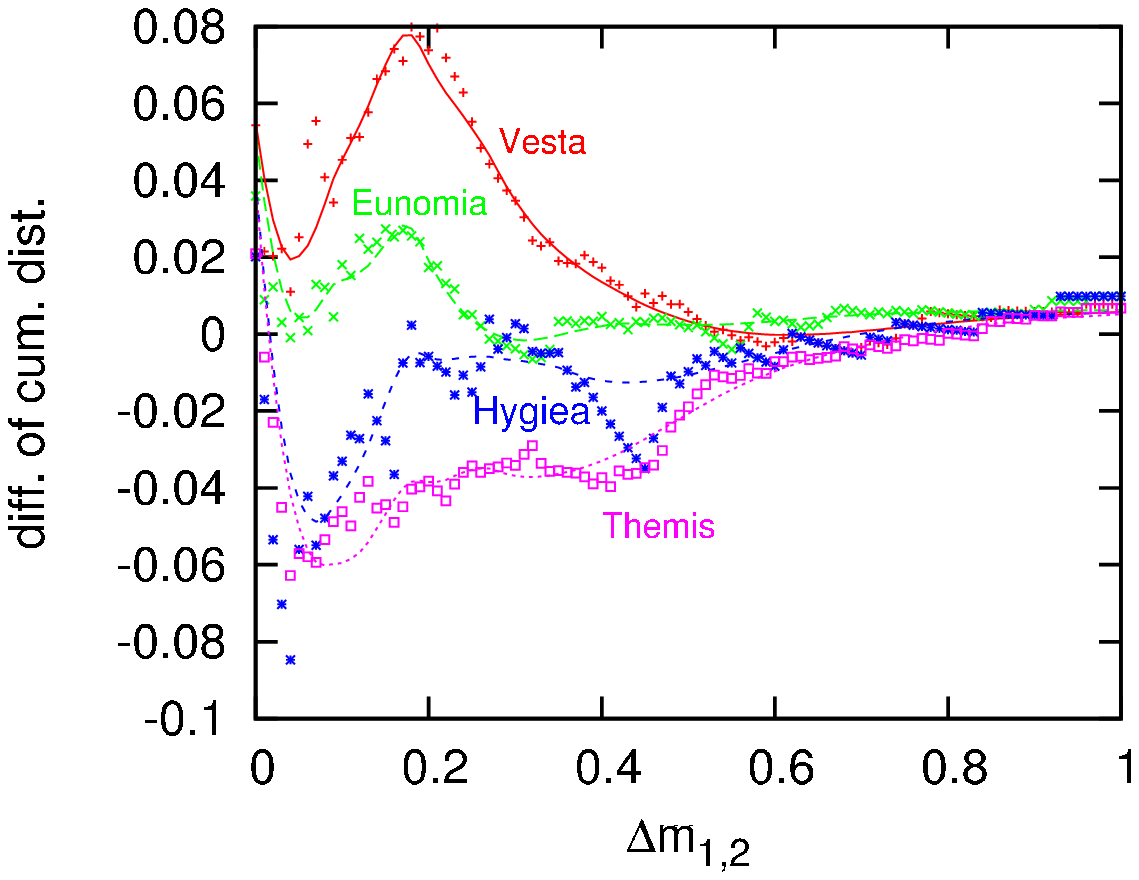}
	\label{allfit}
	\caption{{Upper panels: the cumulative brightness variation of the examined
	families; lower panels: the difference in respect to the cumulative brightness
	variation in the whole sample. Individual symbols show the observational data,
	smooth lines represent the fitted models. The two group of families
	are the same as in Fig. 4.}}
\end{figure}

\newpage

\begin{figure}[h]
	\centering\includegraphics[width=11cm]{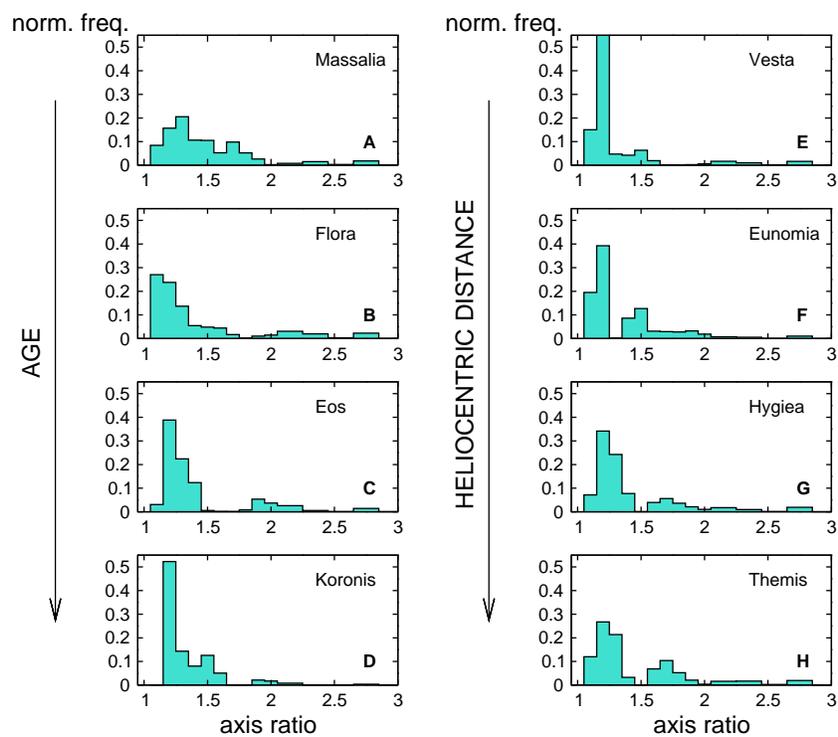}
	\label{allfit}
	\caption{The evolution of shape distribution in asteroid families. A-D: (Massalia to 
	Koronis) families of 150 to 2500 million years show the dependence on age: 
	the more elongated members of young families erode in time toward rough 
	spheroids, whose relative frequency increases up to $50\%$. E-H: (Vesta to 
	Themis) old families at increasing heliocentric distances -- the farther the 
	family, the more elongated its members. Note the distinct peaks of the 
	distributions.}
\end{figure}
\newpage

\begin{figure}[h]
	\centering\includegraphics[width=11cm]{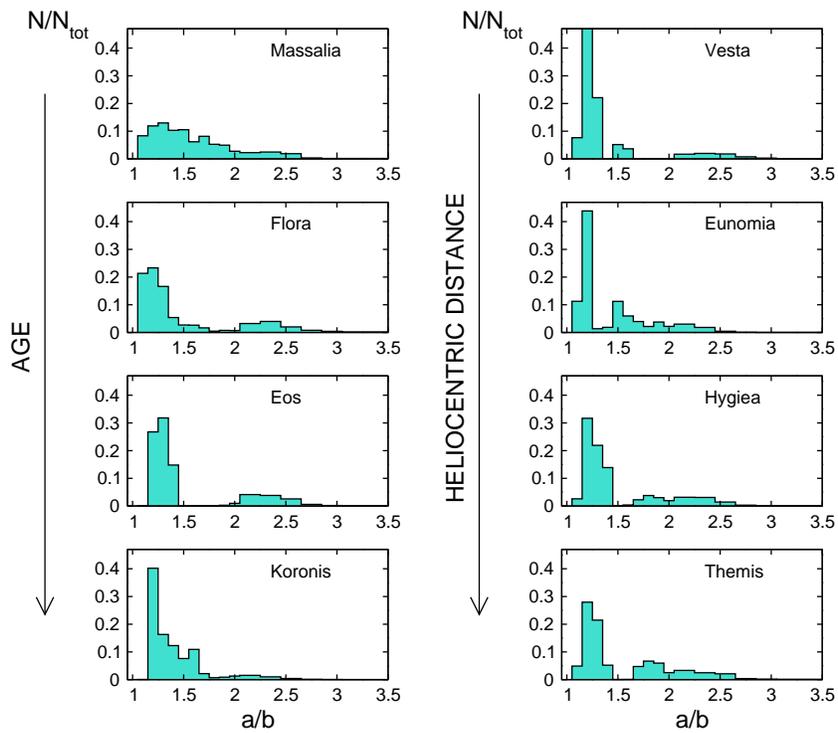}
	\label{appfig}
	\caption{The same as Fig 4, but calculated with model asteroids having randomly orinted spin axis with a fixed latitude of 50 degrees.}
\end{figure}

\newpage

\begin{figure}[h]
	\centering\includegraphics[width=11cm]{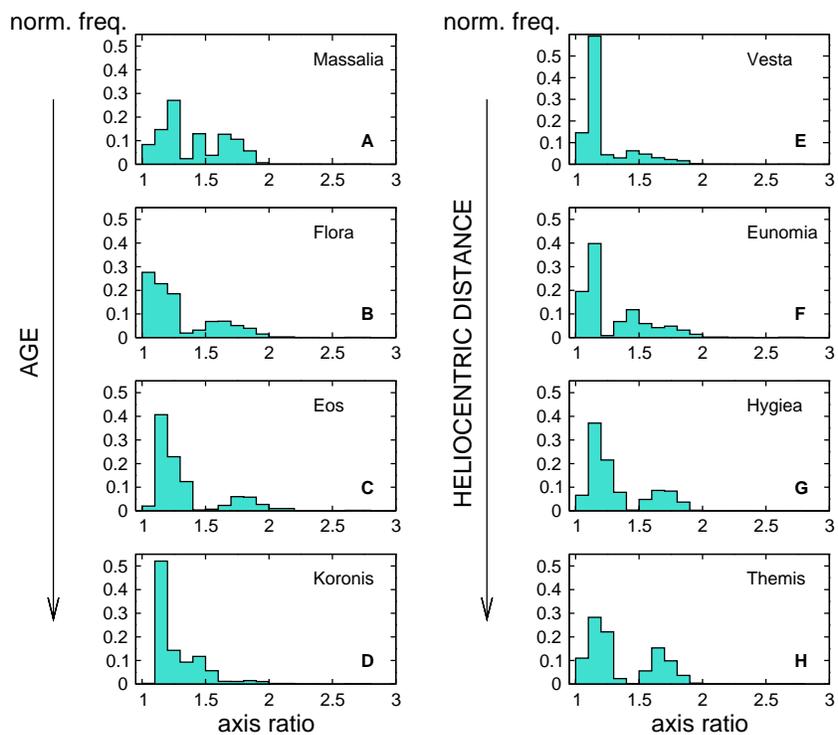}
	\label{appfig}
	\caption{The same as Fig 4, but ignoring asteroids with $\Delta m_{1,2}>0\fm{}65$, with which we checked
	the robustness of the bimodal distributions.}
\end{figure}

\end{document}